# Concept for a Future Super Proton-Proton Collider

Jingyu Tang[1*], J. Scott Berg[13], Weiping Chai[2], Fusan Chen[1], Nian Chen[3], Weiren Chou[1,4], Haiyi Dong[1], Jie Gao[1], Tao Han[5], Yongbin Leng[6], Guangrui Li[7], Ramesh Gupta[13], Peng Li[2], Zhihui Li[8], Baiqi Liu[1], Yudong Liu[1], Xinchou Lou[1], Qing Luo[3], Ernie Malamud[4], Lijun Mao[2], Robert B. Palmer[13], Quanling Peng[1], Yuemei Peng[1], Manqi Ruan1, GianLuca Sabbi[14], Feng Su[1], Shufang Su[9], Diktys Stratakis[13], Baogeng Sun[10], Meifen Wang[1], Jie Wang[3], Liantao Wang[10], Xiangqi Wang[3], Yifang Wang[1], Yong Wang[3], Ming Xiao[1], Qingzhi Xing[7], Qingjin Xu[1], Hongliang Xu[3], Wei Xu[3], Holger Witte[13], Yingbing Yan[6], Yongliang Yang[3], Jiancheng Yang[2], Youjin Yuan[2], Bo Zhang[3], Yuhong Zhang[11], Shuxin Zheng[7], Kun Zhu[12], Zian Zhu[1], Ye Zou[1],

[1]Institute of High Energy Physics, CAS, Beijing 100049, China
[2]Institute of Modern Physics, CAS, China
[3]University of Science and Technology of China
[4]Fermi National Accelerator Laboratory, USA
[5]University of Pittsburgh, USA
[6]Shanghai Institute of Applied Physics, CAS, China
[7]Tsinghua University, Beijing, China
[8]Sichuan University, Chengdu, China
[9]University of Arizona, USA
[10]University of Chicago, USA
[11]Thomas Jefferson National Accelerator Facility, USA
[12]Peking University, China
[13]Brookhaven National Laboratory, USA
[14]Lawrence Berkeley National Laboratory, USA

**Abstract:** Following the discovery of the Higgs boson at LHC, new large colliders are being studied by the international high-energy community to explore Higgs physics in detail and new physics beyond the Standard Model. In China, a two-stage circular collider project CEPC-SPPC is proposed, with the first stage CEPC (Circular Electron Positron Collier, a so-called Higgs factory) focused on Higgs physics, and the second stage SPPC (Super Proton-Proton Collider) focused on new physics beyond the Standard Model. This paper discusses this second stage.

# 1 Introduction

## 1.1 Science reach at the SPPC

SPPC will be an extremely powerful machine, far beyond the scope of the LHC, with center of mass energy 70 TeV, a peak luminosity of 1.2 x $10^{35}$ cm$^{-2}$ s$^{-1}$ (and an integrated luminosity of 30 ab$^{-1}$ assuming 2 interaction points and ten years of running). A later upgrade to even higher luminosities is also possible. It is true that luminosity has a more modest effect on energy reach, in comparison with higher beam energy [1], but raising the luminosity will likely be much cheaper than increasing the energy.

---
[*] Corresponding author: tangjy@ihep.ac.cn

Together the CEPC and SPPC will have the capability to precisely probe Higgs physics. However, what people expect more is that SPPC will explore directly a much larger region of the landscape of new physics models, and make a huge leap in our understanding of the physical world. There are many issues in energy-frontier physics that SPPC will explore, including the mechanism of Electroweak Symmetry Breaking (EWSB) and the nature of the electroweak phase transition, the naturalness problem, and the understanding of dark matter. While these three questions can be correlated, they also point to different exploration directions leading to more fundamental physics principles. SPPC will explore new ground and have great potential for making profound breakthroughs in answering all of these questions.

As a "Higgs factory", the CEPC can measure with high precision the properties of the Higgs boson. With the benchmark integrated luminosity of 5 ab$^{-1}$, a sample of one million Higgs can be obtained and the total Higgs width measured to a relative precision of 2.9%. Using the recoil mass method, CEPC can precisely measure the absolute Higgs couplings to the Z bosons g(HZZ) and the invisible decay branching fraction at the sub percent level, to gluons, W bosons and heavy fermions [g(Hgg), g(HWW), g(Hbb), g(Hcc), and g(H$\tau\tau$)] at the few percentage level. In addition, it can measure the rare decay couplings [g(H$\gamma\gamma$) and g(H$\mu\mu$)] to the 10% level. However, limited by its center of mass energy, CEPC cannot directly measure g(Htt) and g(HHH). These two couplings are extremely important for understanding EWSB and naturalness [2].

Extending the CEPC Higgs factory program, billions of Higgs bosons will be produced at the SPPC. This huge yield will provide important physics opportunities, especially for the rare but relatively clean channels. For example, SPPC can improve the measurement of Higgs-photon coupling, observe the coupling g(H$\mu\mu$), and test other rare decays such as t $\rightarrow$ Hc, H $\rightarrow$$\mu\tau$. Reaching a higher energy threshold than CEPC, SPPC could measure g(HHH) to the 10% level [3], and directly determine the coupling g(Htt) to the sub-percentage level [4]. The Higgs self-coupling is regarded as the holy grail of experimental particle physics, not only because of the experimental challenges, but also because this coupling is a key probe to the form of the Higgs potential. By measuring g(HHH), SPPC can help to answer the question whether the electroweak phase transition is of the 1$^{st}$ order or 2$^{nd}$ order, crucially connected to the idea of electroweak baryogenesis.

As an energy frontier machine, the SPPC could discover an entirely new set of particles in the $O$ (10 TeV) regime, and unveil new fundamental physics principles. One of the most exciting opportunities is to address the naturalness problem. This problem stems from the vast difference between two energy scales: the currently probed electroweak scale and a new fundamental scale, such as the Planck scale. Solutions to the naturalness problem almost inevitably predict the existence of a plethora of new physics particles not far from the electroweak scale. Discovery of such new particles will be a stunning success for an understanding of the naturalness principle. Searching for these possible new particles at the LHC can probe the level of fine-tuning down to $10^{-2}$, while SPPC would push this down to the unprecedented level of $10^{-4}$, beyond the common concept of the naturalness principle.



Dark matter remains one of the most puzzling issues in particle physics and cosmology. Weakly interacting massive particles (WIMPs) are still the most plausible dark matter candidates. If dark matter interacts with Standard Model particles with coupling strength similar to that of the weak interaction, the mass of a WIMP particle could easily be in the TeV range, and likely to be covered at SPPC energy. Combining the relevant bounds on the mass and coupling from the direct (underground) and the indirect (astroparticle) dark matter searches, SPPC would allow us to substantially extend the coverage of the WIMP parameter space for large classes of models.

At the SPPC energy regime, all the SM particles are essentially "massless", and electroweak symmetry and flavor symmetry will be restored. The top quark and electroweak gauge bosons should behave like partons in the initial state, and like narrow jets in the final state. Understanding SM processes in such an unprecedented environment poses new challenges and offers unique opportunities for sharpening our tools in the search for new physics at higher energy scales.

## 1.2 The SPPC Complex

SPPC is a complex accelerator facility and will be able to support research in different fields of physics, similar to the multiuse accelerator complex at CERN. Besides the energy frontier physics program in the collider, the beams from each of the four accelerators in the injector chain can also support their own physics programs. The four stages, shown in Figure 1 and with more details in Figure 8, are a proton linac (p-Linac), a rapid cycling synchrotron (p-RCS), a medium-stage synchrotron (MSS) and the final stage synchrotron (SS). This research can occur during periods when beam is not required by the next-stage accelerator. For example, the high-power proton beam of about 0.8 MW from the p-Linac can be used for production of intense beams of neutrons, muons and rare isotopes for a wide range of research. The high-power beams of 10 GeV from the p-RCS and 180 GeV from the MSS can be used to produce very powerful neutrino beams for neutrino oscillation experiments and the high energy beam from the SS can be used for hadron physics research.

The option of heavy ion collisions also expands the SPPC program into a deeper level of nuclear matter studies. There would also be the possibility of electron-proton and electron ion interactions.

In summary, SPPC will play a central role in experimental particle physics in this post-Higgs discovery world. It is the natural next stage of the circular collider physics program after CEPC. Combining these two world class machines will be a significant milestone in our pursuit of the fundamental laws of nature.

## 1.3 Design goals

SPPC is a proton-proton collider, a discovery machine at the energy frontier. Given the 54.4 km circumference tunnel predefined by CEPC, we will try to achieve the highest possible collision energy in p-p collisions with the anticipated accelerator technology in the 2030's. This, of course, depends on the magnetic field that bends the protons around the ring. Taking into account the



expected evolution in detector technology we can expect that the peak luminosity of $1.2 \times 10^{35}$ cm$^{-2}$s$^{-1}$ will be usable. At least two IPs will be available.

Table 1: Key SPPC parameters

| Parameter | Value | Unit |
|---|---|---|
| Collision energy (C. of M.) | 70.6 | TeV |
| Peak luminosity | $1.2 \times 10^{35}$ | cm$^{-2}$s$^{-1}$ |
| Number of IPs | 2 | |
| Circumference | 54.4 | km |
| Injection energy | 2.1 | TeV |
| Overall cycle time | ~15 | hours |
| Dipole field | 20 | T |

This paper describes what the SPPC will look like, basic design parameters, and its major challenges in accelerator physics and technology. It also explores compatibility in the same tunnel with the previously built CEPC and different operating modes such as electron-proton, proton-ion, and electron-ion. Some key parameters are shown in Table 1.

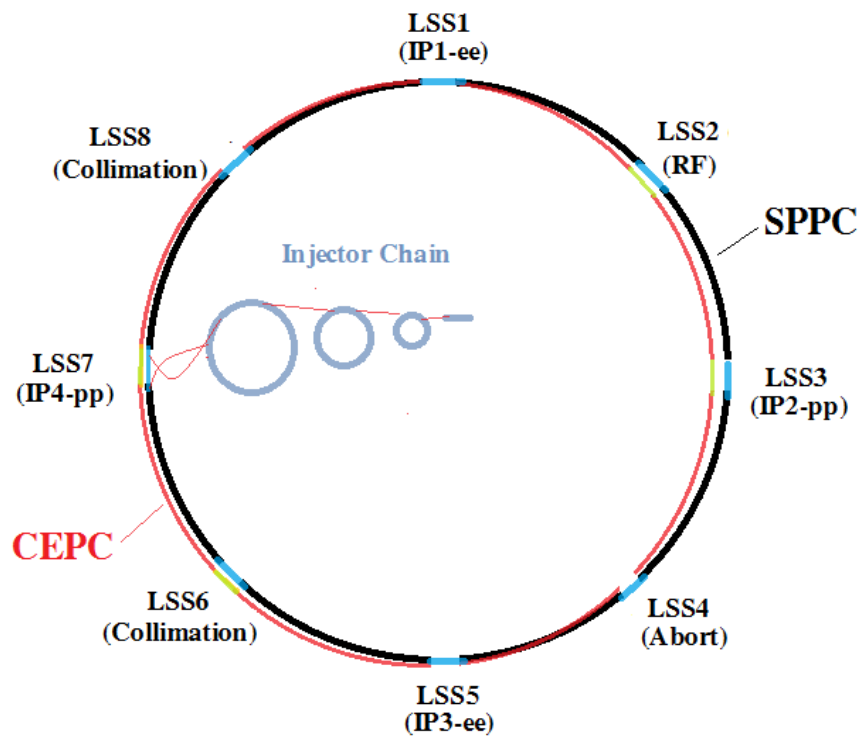

Fig. 1: SPPC accelerator complex

## 1.4 Overview of the SPPC design

The collider will coexist with the previously built CEPC, housed in the same tunnel, of circumference 54.4 km. The shape and symmetry of the tunnel is a compromise between the two



colliders. The SPPC requires relatively longer straight sections which will be described later. This means eight identical arcs, and eight long straight sections for two large detectors, injection and extraction, RF stations and a complicated collimator system. Based on expected progress in high-field magnet technology in the next fifteen to twenty years, we expect that a field of 20 T will be attainable for the main dipole magnets. A hybrid structure of $Nb_3Sn$ and high-temperature superconducting (HTS) conductors with two beam apertures is foreseen. A filling factor of 79% in the arcs (similar to LHC) is assumed. The SPPC will potentially provide beams at a center of mass collision energy of about 70 TeV.

With a circulating beam current of about 1 A and small beta functions (β*) of 0.75 m at the collision points, the peak luminosity can reach $1.2 \times 10^{35}$ cm$^{-2}$s$^{-1}$. The high beam energy, high beam current and high magnetic field will produce very strong synchrotron radiation which will impose critical requirements on the vacuum system which is based on cryogenic pumping. We expect that this technical challenge will be solved in the next two decades by developing efficient beam screens to extract the heavy heat load from the synchrotron radiation and reduce the electron cloud density within the beam apertures. If forced to lower the synchrotron radiation power, we would have to reduce the bunch population or the number of bunches and try to achieve a smaller β*.

As in other proton colliders using superconducting magnets, the injection energy is mainly defined by the field quality of the magnets at the bottom of their range. Persistent currents in the coils (magnetization) distort the field distribution at injection energy. Other factors favoring relatively higher injection energy are the coupling impedance, which is important to collective beam instabilities; the smaller emittance required to reduce apertures of beam screen and magnet, and the requirement on the good-field-region of the magnets. If we use the LHC ratio of 15 for top to bottom fields the injection energy would be 2.37 TeV. A larger ratio of 20 could be considered, which would mean an injection energy of 1.78 TeV. This would make the injector chain cheaper. In this report, we have adopted a compromise with an injection energy of 2.1 TeV.

The injector chain pre-accelerates the beam to injection energy with the required bunch current, bunch structure, and emittance. The injection chain determines the beam fill period. To reach 2.1 TeV, a four-stage injector chain is proposed: the p-Linac to 1.2 GeV, the p-RCS to 10 GeV, the MSS to 180 GeV and the SS to 2.1 TeV. High repetition rates for the lower energy stages help reduce the SS cycling period. This is important because the SS uses superconducting magnets. The beams of high repetition rates can also be used for other research applications when the accelerators are not preparing beam for injection into the SPPC.

If not controlled, synchrotron cooling would rapidly reduce the beam emittances and cause excessive beam-beam tune shifts. Noise in transverse deflecting cavities must be used to limit the minimum transverse emittances, and thus tune shifts. Without leveling, and with constant beam-beam tune shift, the luminosity decays exponentially from its peak with a lifetime of approximately 10 hours. To maximize the integrated luminosity, the turnaround time (defined as the period in a machine cycle excluding the collision period) should be made as short as possible, preferably short compared to the beam decay time. The initially assumed average 3-hour is acceptable, giving an optimized complete cycle time of about 10 hours, but a turnaround time of as little as 0.77 hour s would certainly be preferred.



The peak and average luminosities could be raised by allowing the synchrotron damping to lower the transverse emittance and allowing higher but acceptable tune shifts (0.02-0.03). But, if not leveled, the peak luminosities and thus the numbers of interactions per beam crossing could become excessive. Limiting the peak luminosity (leveling) would limit this number, yet still allow an increase in the average luminosity. Using more and closer spaced bunches could reduce the number of interactions per bunch crossing, without lowering the peak luminosities. However, if the beam current is not to be raised, the numbers of protons per bunch must be proportionally reduced, and, if luminosity is to be preserved, the synchrotron damping must be allowed to further lower the emittances, while not increasing the tune shifts. Whether closer bunch spacing is consistent with electron cloud considerations is yet to be determined.

Lowering the beta functions at the collision points could further increase luminosities without increasing the tune shift. If this was done after the emittances have been damped, then larger aperture final triplet magnets, or requiring them to be closer to the IP, are not required. This option will be studied.

There are many technical challenges in designing and building the collider, including its injector chain. The two most difficult are the development and production of 20-T magnets, and the beam screen associated with very strong synchrotron radiation. Significant R&D efforts in the coming decade are needed to solve these problems.

## 2 Key accelerator physics issues

## 2.1 Main parameters

### 2.1.1 Collision energy and layout

To reach the design goal for the 70 TeV center of mass energy with this relatively small circumference of 54.4 km, very high-field magnets of about 20 T have to be used. A hybrid structure of $Nb_3Sn$ and High Temperature Superconducting (HTS) conductors will be used. In addition, the ring should be designed to be as compact as possible. Although the lattice has not been designed, it is assumed to be a traditional FODO everywhere, except at the IPs where triplets are used to produce the very small $\beta*$. One can make a very preliminary outline design for the SPPC without a real lattice. The arcs represent most of the circumference, and the arc filling factor is taken as 0.79, similar to LHC [5]. A key issue here is to define the number of long straight sections and their lengths. They are needed to produce those very small beta functions where the large physics detectors are to be placed, and for hosting the beam injection and extraction systems (abort), collimation systems and RF stations. Some compromises have to be made to have a relatively compact design of the long straight sections. Our design is more compact than LHC, and is compatible with the CEPC layout. A total length of about 7.6 km is reserved for the long straight sections, with eight long straight sections of which 4 are 1100 m long and the 4 others are 850 m long. With this configuration, the top beam energy is 35.3 TeV which provides 70.6 TeV in collision energy. The main parameters are listed in Table 2.



Table 2: Main SPPC parameters

| Parameter | Value | Unit |
|---|---|---|
| **Main parameters** | | |
| Circumference | 54.4 | km |
| Beam energy | 35.3 | TeV |
| Lorentz gamma | 37644 | |
| Dipole field | 20 | T |
| Dipole curvature radius | 5885 | m |
| Arc filling factor | 0.79 | |
| Total dipole magnet length | 36977 | m |
| Arc length | 46806 | m |
| Total straight section length | 7554 | m |
| Energy gain factor in collider rings | 16.8 | |
| Injection energy | 2.1 | TeV |
| Number of IPs | 2 | |
| Revolution frequency | 5.52 | kHz |
| **Physics performance and beam parameters** | | |
| Peak luminosity per IP | $1.2 \times 10^{35}$ | $cm^{-2}s^{-1}$ |
| Beta function at collision | 0.75 | m |
| Circulating beam current | 1.0 | A |
| Nominal beam-beam tune shift limit per IP | 0.006 | |
| Bunch separation | 25 | ns |
| Number of bunches | 5798 | |
| Bunch population | $2.0 \times 10^{11}$ | |
| Accumulated particles per beam | $1.2 \times 10^{15}$ | |
| Normalized rms transverse emittance | 4.1 | μm |
| Beam life time due to burn-off | 9.6 | hours |
| Total inelastic cross section | 140 | mb |
| Reduction factor in luminosity | 0.96 | |
| Full crossing angle | 73 | μrad |
| rms bunch length | 75.5 | mm |
| rms IP spot size | 9.0 | μm |
| Beta at the first parasitic encounter | 19.5 | m |
| rms spot size at the first parasitic encounter | 45.9 | μm |
| Stored energy per beam | 6.6 | GJ |
| SR power per beam | 2.1 | MW |
| SR heat load at arc dipoles | 56.9 | W/m |
| Energy loss per turn | 2.06 | MeV |



### 2.1.2 Luminosity

The initial luminosity of $1.2 \times 10^{35}$ cm$^{-2}$s$^{-1}$ is much higher than in previously built machines such as the Tevatron [6] and LHC [5] and in designs such as SSC [7], VLHC [8], HE-LHC [9], and more than a factor of two higher than FCC-hh [10], though perhaps lower than in the HL-LHC [11]. In order to achieve this high luminosity, a large number of bunches and high bunch population are needed. These will be supplied by a powerful injector chain.

The SPPC initial luminosity being approximately 2.5 times higher than the FCC-hh [10], while using the same bunch spacing, the number of interactions per bunch crossing is higher than present-day detectors could handle. It is believed, however, that ongoing R&D efforts on detectors and general technical evolution will be able to solve this problem. It also requires double the number of protons per bunch of the FCC-hh, double the current, and a somewhat smaller β*.

Besides the challenges in the detectors, very high synchrotron radiation and very strict beam loss control associated with the high circulation current of 1 A are major challenges to the vacuum system and the machine protection system.

Another important parameter is the average, and thus integrated luminosity. One must consider the loss of stored protons from collisions, the cycle turnaround time, and the shrinking of the transverse emittance due to synchrotron radiation. Beam decay and turnaround time reduce the integrated luminosity. Emittance shrinkage from synchrotron radiation could maintain or even raise the peak luminosity after the collision start, but would eventually also increase the beam-beam tune shift to an unacceptable level. An emittance blow-up system is thus used to counteract the emittance shrinkage, and can be used to limit the tune shift to an acceptable level. Another method to increase the luminosity is to adjust β* during the collisions by taking advantage of emittance shrinking while keeping the beam-beam tune shift constant.

### 2.1.3 Bunch structure and population

Many bunches with relatively small bunch spacing are desirable for achieving high luminosity operation. However, the bunch spacing can be limited both by parasitic collisions in the proximity of the IPs, and by the electron cloud instability. One also needs to consider the ability of the detector trigger systems to cope with short bunch spacing. Although the bunch gap of 25 ns was designed as a baseline for LHC, the machine has been operating to date with 50-ns bunch spacing. This was due to problems in operation mainly from the electron cloud effect. It is believed that the problems related to 25 ns at LHC will be overcome in the near future. Therefore, we have also adopted 25 ns for the nominal bunch spacing at SPPC. The bunch spacing of 25 ns is defined by the RF system in the MSS of the injector chain and preserved from there on. The possibility of shorter bunch spacing will be investigated and is discussed below in Section 2.2.

Time gaps between multiple bunch trains are needed for beam injection and extraction in both SPPC and the injector chain. Their lengths depend on the practical designs of the injection and extraction (abort) systems, and the rise time of the kickers for beam extraction from SPPC, assumed now to be a few microseconds. The bunch filling is taken to be about 80% of the ring circumference, similar to LHC. These gaps in the bunch structure have a significant impact on the beam dynamics during collision. On the one hand, the gaps between bunch trains are useful



in suppressing collective beam instabilities; on the other hand, they give different average numbers of collisions per revolution for different bunches, and this will produce differing beam-beam effects.

Bunch population is first defined in the p-RCS of the injector chain, where the beam from the p-Linac fills the RF buckets using both transverse and longitudinal paintings. Similar to that in the SPL linac for LHC, with a relatively high-energy linac beam, one can obtain a high bunch population, with acceptable space charge effects in the p-RCS. Each long bunch from the p-RCS will be split evenly into many smaller bunches in the MSS, using an RF system that defines the nominal bunch population and spacing. With the nominal bunch number and bunch population, the circulation current will be about 1 A in the collider rings, similar to that of the future HL-LHC [11] and double that of FCC-hh [10].

### 2.1.4 Beam sizes at the IPs

The beam sizes are determined by the β* of the insertion lattice and the beam emittance. The initial normalized emittance is predefined in the p-RCS of the injector chain and preserved with a slight increase in the course of reaching the top energy of the SPPC due to many different factors such as nonlinear resonance crossings. However, at the maximum energy of 35.3 TeV and in the later part of the acceleration stage, synchrotron radiation will take effect, with damping times about 1.0 hours and 0.5 hours for the transverse and longitudinal emittances, respectively. This will allow emittances after the start, significantly smaller than their initial values. However, the emittances cannot be allowed to fall without limit because of beam-beam tune shift and luminosity considerations. Thus a stochastic emittance heating system is required to limit the synchrotron radiation cooling and control the emittance level during the collision process.

### 2.1.6 Crossing angle at the IPs

To avoid parasitic collisions near the IPs producing background for experiments, it is important to separate the two beams, except at the IPs, using a crossing angle between the two beams. The crossing angle is chosen to avoid the beams overlapping at the first parasitic encounters at 7.5 m from the IPs when the bunch spacing is 25 ns. At these locations the separation is no less than 12 times the rms beam size. At the SPPC, this crossing angle at the collision energy is about 75 μrad. Compared with head-on collisions, this bunch crossing angle will result in a few percent luminosity reduction. The crossing angle could be increased later in a more realistic design, and would have to be increased if smaller bunch spacing were to be adopted, as discussed in Section 2.2.

With a small bunch separation the crossing angle must be larger, and this reduction of luminosity would be larger if not controlled with crab cavities. There is no luminosity loss with crossing angles when crab cavities are used. The crossing angle may be different at injection due to different lattice settings and larger emittance.

At the superconducting quadrupole triplets, the two beams are separated from each other by the crossing angle, and the apertures of the quadrupoles are increased significantly.



### 2.1.7 Turnaround time

Turnaround time is the total time period in a machine cycle when the beams are out of collision, including the programmed count down checking time before injection, the final check with a pilot shot, the beam filling time with SS beam pulses, the ramping up (or acceleration) time, and the ramping down time. Filling one SPPC ring requires 6 SS beam pulses, which means a minimum filling time of about 5 minutes including pilot pulses. The ramping up and down times are each about 18 minutes. Altogether, the minimum turnaround time is 46 minutes, or about 0.77 hour. However, the experience at LHC and other proton colliders [12] shows that only about one third of the operations from injection to the top energy are successful and the average turnaround time is closer to 3 hours. This is considered acceptable, and with a luminosity run time of 4-8 hours, during which the beams are in collision, it gives a total cycle time of about 7-11 hours.

### 2.1.8 RF parameters

The main acceleration system at SPPC uses 400-MHz superconducting cavities. However, an additional RF system of 200 MHz is considered helpful for the longitudinal matching from the SS to the collider during injection. Although the ramping-up time is mainly defined by the superconducting magnets, the RF system must provide sufficient voltage during the process to keep up the acceleration rate with a large longitudinal acceptance. When nearing the final stage of acceleration, synchrotron radiation will play a significant role. About 10 MV in RF voltage is needed to compensate the synchrotron radiation, and the situation is similar during the collisions (and the preparation phase bringing the beams into collision). A total RF voltage of either 24 or 32 MV per beam will be provided by the 400-MHz system. Stochastic noise must be introduced to raise the longitudinal emittance to give the long bunches required to avoid detector pile up, and avoid instabilities.

## 2.2 Synchrotron radiation

Synchrotron radiation (SR) power is proportional to the fourth power of the Lorentz factor and the inverse of the radius of curvature in the dipoles, and becomes an important effect at multi-TeV energies using high field superconducting dipoles. With the beam current of 1 A and magnetic field of 20 T, the synchrotron radiation power reaches about 57 W/m per aperture in the arc sections, more than two orders higher than that at LHC. The average critical photon energy is about 2.1 keV. There is also a synchrotron radiation effect in the high-gradient superconducting quadrupole magnets. The technical challenges of the vacuum system and beam screen are discussed in Section 3.2.

At the SPPC, synchrotron radiation imposes severe technical challenges to the vacuum system and a probable limit on the circulating current. If absorbed at the liquid helium temperature of the magnet bores, the synchrotron radiation's heat load would be excessive, so it must be absorbed at a higher temperature. A beam screen, or other capture system, must be situated between the beam and the vacuum chamber. This limits the beam tube aperture, raising the beam impedance, and/or increases the required superconducting magnet bore radius. The



working temperature at the beam screen is a key parameter in the design. The beam screen is also important in controlling the coupling impedance and reducing the electron cloud effect.

If the synchrotron radiation falls directly on the inside of the beam screen then it will propagate far along the pipe with multiple very small angle reflections. It then becomes distributed around the bore of the pipe. The photo-electrons generated feed the electron cloud. Allowing the photons to pass through a slit in the screen and, and then be deflected into photon absorption channels, as discussed for the FCC-hh [13], could be an effective way to reduce this problem.

Table 3: Relevant parameters during operation with bunch spacing of 25 ns and: (a) a fixed tune shift; (b) allowing the tune shift to rise to 0.03; (c) as in (b) but levelling the luminosity to its initial value; (d) as for (c) but with bunch spacing of 10 ns; (e) as for (d) but reducing β* proportional to the emittance down to 25 cm; (f) as for (e) but with bunch spacing of 5 ns. All values are for run times maximized for a turnaround time of 3 hr., except for the parenthesized average luminosities that are for turnaround times of 0.77 hr.

| | Collis. time | Bunch spacing | Events/ crossing | Luminosity | Norm. emittance | Protons/ bunch | Tune shift | Beta* |
|---|---|---|---|---|---|---|---|---|
| | hours | ns | | $10^{35}\,\mathrm{cm^{-2}s^{-1}}$ | mm-mrad | $10^{11}$ | | cm |
| (a) | 6.9 | 25 | 490 | Max 1.24 | Init. 4.1 | Init. 2.0 | Init. 0.01 | Init. 75 |
| | | | | Ave 0.59 (0.81) | Final 2.20 | Final 1.01 | Final 0.01 | Final 75 |
| (b) | 4.25 | 25 | 1120 | Max 2.85 | Init. 4.1 | Init. 2.0 | Init. 0.01 | Init. 75 |
| | | | | Ave 1.11 (1.69) | Final 0.64 | Final 0.51 | Final 0.03 | Final 75 |
| (c) | 8.0 | 25 | 490 | Max 1.24 | Init. 4.1 | Init. 2.0 | Init. 0.01 | Init. 75 |
| | | | | Ave 0.86 (1.11) | Final 0.30 | Final 0.49 | Final 0.03 | Final 75 |
| (d) | 5.2 | 10 | 415 | Max 2.64 | Init. 4.1 | Init. 2.0 | Init. 0.01 | Init. 75 |
| | | | | Ave 1.02 (1.45) | Final 0.17 | Final 0.24 | Final 0.03 | Final 75 |
| (e) | 4.0 | 10 | 490 | Max 3.10 | Init. 4.1 | Init. 2.0 | Init. 0.01 | Init.75 |
| | | | | Ave 1.43 (2.12) | Final 0.16 | Final 0.13 | Final 0.03 | Final 25 |
| (f) | 3.9 | 5 | 300 | Max 3.93 | Init. 4.1 | Init. 2.0 | Init. 0.01 | Init. 75 |
| | | | | Ave 1.40 (2.11) | Final 0.12 | Final 0.08 | Final 0.03 | Final 25 |

The synchrotron radiation also has an important impact on the beam dynamics at, and approaching, the top energy. Without intervention, both the longitudinal and transverse emittances will shrink with lifetimes of about 0.5 and 1.0 hours, respectively. The short damping times may help eliminate collective beam instabilities. One may exploit this feature to enhance the machine performance by allowing the transverse emittances to fall and to increase the



luminosity. But nevertheless, to avoid excessive beam-beam tune shift (see Section 2.3), a stochastic transverse field noise systems will have to be installed to control the emittance reduction.

Table 3 and Figure 2 show the relevant parameters as a function of time for six cases. In all cases, the collision times are chosen to give the maximum average luminosities assuming the baseline turnaround time of 3 hours. The increased average luminosities with an ideal turnaround time of 0.77 hours are shown in parentheses in Column 5.

Case (a) assumes that the transverse stochastic noise is adjusted to keep the beam-beam tune shift to its initial value of 0.01. The bunch population, emittance and luminosity, all fall exponentially with time. The peak luminosity is its initial value of $1.24 \times 10^{35} \mathrm{cm}^{-2}\mathrm{s}^{-1}$ giving 490 events per bunch crossing, which is considered manageable. The average luminosity is $0.59 \times 10^{35} \mathrm{cm}^{-2}\mathrm{s}^{-1}$, only about half its initial value.

In Case (b), the noise is reduced to allow the beam-beam tune shift to rise to 0.03, but then modified to keep it at that value. The average luminosity is now $1.11 \times 10^{35} \mathrm{cm}^{-2}\mathrm{s}^{-1}$, almost equal to its initial value, and this is a considerable gain. But the peak luminosity is now $2.2 \times 10^{35} \mathrm{cm}^{-2}\mathrm{s}^{-1}$, giving 1120 events per bunch crossing which is excessive.

Case (c), is the same as Case (a) but adds the constraint of keeping the peak luminosity no higher than its initial value, corresponding to 490 events per bunch crossing (see Section 2.1.6). The average luminosity is now down to $0.86 \times 10^{35} \mathrm{cm}^{-2}\mathrm{s}^{-1}$ or 69% of its initial value, but still significantly better than Case (a).

Case (d) is the same as Case (c) but the bunch spacing has been reduced from 25 to 10 ns, and the bunch intensity is decreased by the same factor of 2.5 from $2 \times 10^{11}$ to $4 \times 10^{10}$, to keep the circulation current constant. This lowers the initial luminosity by the same factor of 2.5, but increases the average luminosity to $1.02 \times 10^{35} \mathrm{cm}^{-2}\mathrm{s}^{-1}$, and the peak luminosity as well. However, now with higher bunch frequency, the maximum number of events per bunch crossing has been reduced to 415 that is less than our assumed limit of 490, so no luminosity leveling is required.

Case (e) is the same as Case (d) but with dynamic $\beta^*$ reduction (see Section 2.1.7), in which, as the transverse emittance falls, the $\beta^*$ is reduced in proportion, until it reaches 25 cm. Luminosity leveling is now required to reduce the maximum luminosity, but the average luminosity is still increased to $1.43 \times 10^{35} \mathrm{cm}^{-2}\mathrm{s}^{-1}$ which is more than twice that of the conservative Case (a).

Case (f) is the same as Case (e) but now with bunch spacing of only 5 ns. It has almost the same average luminosity as Case (e), but the peak luminosity is lower, with the maximum event per bunch crossing only 300.

In the "Luminosity" column of Table 3, in parentheses, the average luminosities are also given for an ideal turnaround time of 0.77 hours, showing that further improvements on the average luminosities are significant. Such a short turnaround time may need a full circumference accumulator. The shorter bunch spacings in Cases (d), (e), and (f) might also require an upgrades to the injection chain. Electron cloud instabilities (Section 2.4) with the shorter bunch spacing need more study, and may favor either 25 ns or 5 ns, rather than the intermediate spacing of 10 ns.



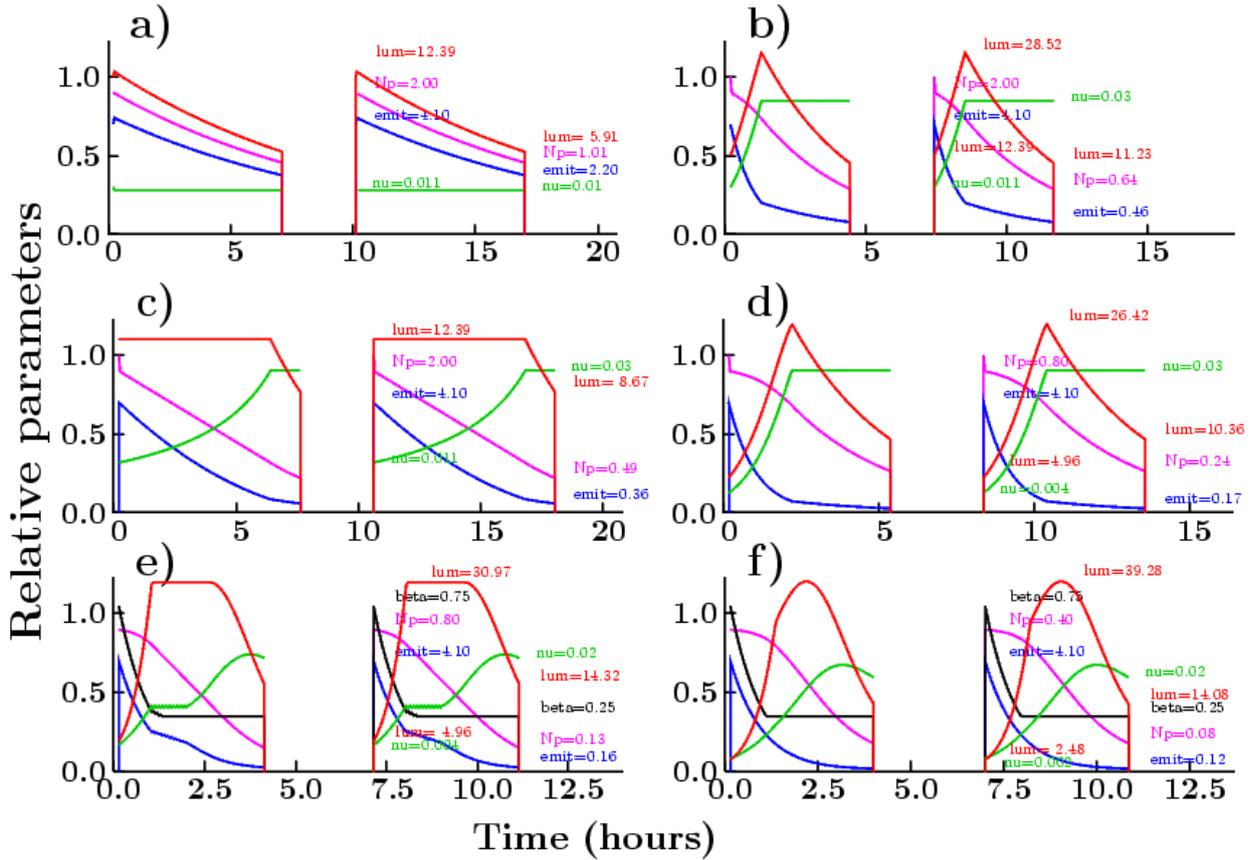

Fig. 2: Evolution of parameters vs time with a turnaround time of 3 hours and bunch spacing of 25 ns. Red: luminosity, magenta: number of protons per bunch, blue: transverse emittance, green: beam-beam tune shift, black: beta* at the IP. (a) with fixed tune shift; (b) allowing the tune shift to rise to 0.03; (c) as in (b) but with the luminosity "leveled" at its initial value; (d) as in (c) but bunch spacing of 10 ns; (e) as for (d) but reducing beta* in proportion to emittance down to 25 cm; (f) as for (e) but with bunch spacing of 5 ns.

.

### 2.2.1 Intra-beam-scattering IBS

Intra-beam scattering within bunches can couple longitudinal momentum into transverse motion, and it will increase the transverse emittances, or in our case, slow the emittance cooling from synchrotron radiation. With the initial parameters, IBS has a negligible effect. But as the emittance shrinks from the synchrotron cooling, it becomes significant, and eventually limits the emittance reduction. For the evolution calculations in Table 3 and Figure 2, it was included, using the approximate scaling rule:

$$\tau_{IBS}(hr.) \approx 0.28 \, \frac{\sqrt{E_p(TeV)} \, \sigma_z(m) \, C^2(km)}{n_p(10^{10})} \left( \frac{\epsilon(\mu m)}{<\beta>(km)} \right)^{2.5}$$

where $E_p$ is the beam energy, $\sigma_z$ is the rms bunch length, $C$ is the ring circumference, $n_p$ is the number of protons per bunch, $\varepsilon$ is the transverse normalize emittance, and $<\beta>$ is an estimate of the average beta in the arcs, taken to be 248 m.

### 2.2.2 Luminosity leveling

As noted above, even with the initial luminosities, the numbers of events per bunch crossing (490) is higher than current detectors could accept, but it is assumed to be acceptable with future detector technology. However, as noted in Section 2.2, and shown in Figure 2b, synchrotron cooling of the transverse emittance can generate luminosities greater than its initial value, and further raise the numbers of events per bunch crossing. Optimum physics use would then require a constraint on the events per bunch crossing, requiring a mechanism to limit the maximum luminosities. Such 'luminosity leveling' could be achieved by control of either the $\beta^*$ or the emittances using the stochastic heating system.

### 2.1.7 Dynamic β* reduction

To avoid beam loss, the beam rms size $\sigma$ must be kept below a given minimum fraction of the triplet apertures at the IPs. If $L^*$ is the distance from the triplet to the IP, then the beam size there is given approximately by $\sigma \sim L^* \varepsilon / \beta^*$, which sets a minimum acceptable $\beta^*$. However, as the emittance $\varepsilon$ falls from synchrotron damping, then the $\beta^*$ can be reduced in proportion, without increasing $\sigma$. A lower limit for $\beta^*$ may be set by lattice considerations, and it should not approach too close to the bunch length to avoid hour-glass effects. In the examples in Table 3 and Figure 2, $\beta^*$ reduction was limited to 25 cm, one third of its initial value of 75 cm.

### 2.3 Beam-beam effects

Beam-beam effects, which could lead to emittance growth, lifetime drop, and instabilities, have a very important effect on the luminosity of a collider. There are several different beam-beam effects affecting the performance of a proton-proton collider: the incoherent beam-beam effects which influence beam lifetime and dynamic aperture; the PACMAN effects which will cause bunch to bunch variation; and coherent effects which will lead to beam oscillations and instabilities.

The nominal parameters given in Table 2 are used for the preliminary study of beam-beam effects. By using the beam-beam theory in the reference [14], one obtains an estimate for the beam-beam limit $\xi_{y,max}$=0.0064 per IP. It is reasonable to choose a nominal conservative beam-beam parameter as 0.006. However LHC has reported stable operation with a total value of $\Delta Q_{tot} \sim 0.03$ with 3 interaction points [15], so this limit was used for the examples in Figure 2.

### 2.3.1 Incoherent effects

Each particle in a beam will feel a strong nonlinear force when the beam encounters the counter rotating beam, with deleterious effects on the dynamic behavior of the particle. This nonlinear interaction will lead to an amplitude dependent tune spread for the particles in both transverse planes, which should be studied to keep the tunes away from crossing dangerous resonance lines. Earlier experiences at both the Tevatron [6] and LHC [5], required the total tune spread from all IP crossings to be kept to no more than 0.015. As an example, a beam-beam tune footprint [16] with 2 head-on interactions at SPPC (using the LHC tunes) is shown in Fig. 3. From the plot one can see that the footprint at small amplitudes is crossed by 10th order and 11th order resonances and at higher amplitudes by 13th order resonances. Thus, dynamic aperture is reduced by the beam-beam interaction at the IP, which may lead to beam loss. Therefore, tunes slightly above the LHC values would seem to be a conservative choice.

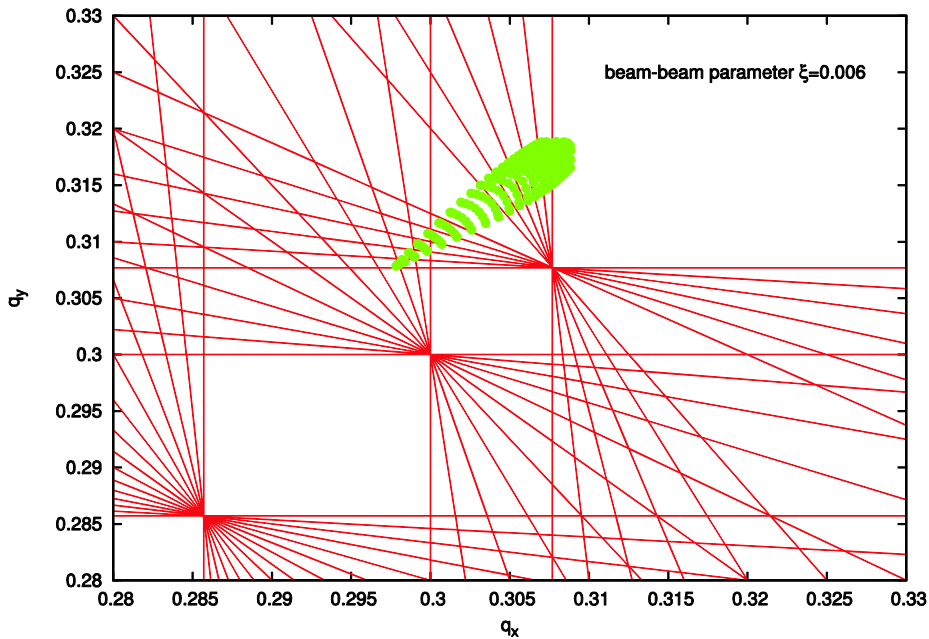

Fig. 3: Two dimensional tune distribution versus amplitude (footprint)

### 2.3.2 PACMAN effects

The circumference and bunch number at SPPC are both about twice those at LHC. With the similar bunch spacing of 25 ns it is expected that the PACMAN effects may have a similar influence as seen at LHC. Only about half of the bunches at SPPC would be regular bunches. The identification of regular bunches is important since the measurements such as tune, orbit or chromaticity should be selectively performed on those. We have to choose a proper fill pattern and crossing scheme to reduce these effects.

### 2.3.3 Coherent effects

Coherent beam-beam effects would be expected in SPPC because the two colliding beams are equally strong. Coherent modes of oscillations of the two counter rotating beams are coupled by the beam-beam interaction; the coherent dipole mode is the most dangerous mode where a bunch oscillates as a rigid object around its nominal orbit. According to LHC experience, it might be an option to use asymmetric collisions (different bunch intensities) at SPPC to suppress the excitation of the coherent mode due to the beam-beam effect.

### 2.3.4 Beam-beam tune shift limits

In order to achieve a higher luminosity, new ideas and technologies are under study, such as the crab waist collision scheme, beam feedback, and other ideas. They could be effective for increasing collider luminosity. New theory and simulation work could guide the study for a luminosity upgrade in the future. The beam-beam simulations by Ohmi [17] predict that the beam-beam limit at LHC might be even larger than the observed maximum of 0.03. By including SR emittance shrinkage and proton burn-off, it is hoped to achieve a much higher integrated luminosity by this method.

## 2.4 Electron cloud effect

The electron cloud (EC) can cause beam instability. The build-up of accumulated photon electrons and secondary electrons has proved to be one of the most serious restrictions on collider luminosity in PEP II, KEKB, LHC [18-19], and BEPC. The EC links together the motion of subsequent bunches and induces coupled bunch instability. It also leads to emittance blow-up and luminosity degradation [20-21]. For next-generation super proton colliders such as SPPC, a bunch population higher than $10^{11}$ and a bunch spacing less than or equal to 25 ns, the EC effect will be critical for reaching the luminosity level of $1.2 \times 10^{35}$ cm$^{-2}$s$^{-1}$.

There are three sources for the electron cloud: photon electrons, residual gas ionization and secondary electron emission. At a vacuum of about 1.0 nTorr, the residual gas density is about $2 \times 10^{13}$ m$^{-3}$. With an ionization cross section of 2.0 Mb, the electrons produced by gas ionization can be ignored. The necessary condition for electron amplification is that the average secondary electron emission yield (SEY) exceeds one. Electron multipacting occurs if the electrons emitted from the wall reach the opposite side wall just prior to the arrival of the next bunch. The criterion $n = \frac{r^2}{n_b r_e L_{sep}}$ can be used to estimate which kind of electrons are the dominant component in the electron cloud. In the formula, $r$ is the radius of the vacuum pipe, $n_b$ the number of particles in the bunch, $L_{sep}$ is the bunch spacing and $r_e$=2.8$\times$10$^{-15}$ m, the classical electron radius. If $n$<1, some of the primary electrons are lost before the next bunch arrives and secondary electrons dominate the electron cloud; if $n$>1, the primary electrons interact with more than one bunch and photon electrons compose most of the electron cloud. The estimated parameter $n$ for different pp colliders are listed in Table 4. The EC build-up saturates when the attractive beam field at the chamber wall is compensated on the average by the electron space charge field. The line density of the electron cloud in the vacuum chamber is $\lambda_e = n_b/L_{sep}$, which corresponds to the volume density $\rho_{e,neutr} \approx \frac{\lambda_e}{\pi ab}$, where $a$ and $b$ are half sizes of the elliptical vacuum pipe. According to the estimated neutralization density shown in Table 4, the EC density in the SPPC rings will be comparable to those at LHC and FCC-hh.

Table 4: Estimates on electron cloud instability for some super pp colliders [2, 22]

| | LHC | FCC-hh | SPPC |
|---|---|---|---|
| Bunch particles ($10^{11}$) | 1.15 | 1.0 | 0.4/0.8/2.0 |
| Bunch spacing (ns) | 25 | 25 | 5/10/25 |
| Beam energy (TeV) | 7 | 50 | 31.7 |
| Pipe radius (mm) | 20 | 13 | 20 |
| Parameter $n$ | 0.165 | 0.189 | 2.37/0.59/0.095 |
| Neutralization line density ($10^{10}$/m) | 1.53 | 1.33 | 2.66 |
| Neutralization volume density ($10^{13}$/m$^3$) | 1.22 | 2.51 | 2.12 |
| Wake field $W/L$ ($10^3$/m$^2$) | 1.33 | 3.15 | 1.33 |
| Betatron tune | 43.3 | - | 60.3 |
| Synchrotron tune | 0.006 | 0.002 | 0.005 |
| Growth time (ms) | 4.31 | - | 4.15 |
| Circumference (km) | 26.7 | 100 | 50 |
| Threshold electron density ($10^{13}$/m$^3$) | 0.66 | 0.147 | 0.468 |

The EC links oscillation between subsequent bunches and may lead to coupled bunch instability. The action propagated by the EC between subsequent bunches can be presented as a wake field expressed as $W_{ec,x,y}/L = 4\pi\rho_{e,neutr}/N_b$, which gives the dipole component per unit length of the wake field. Based on the wake field, the growth rate for the coupled bunch instability is $\frac{1}{\tau_{e,CB}} = \frac{2r_p N_b c^2}{\gamma\omega_\beta ab L_{sep}}$. The coupled bunch instability can be damped by a feedback system. The EC also drives transverse emittance blow-up, which is very important at lower energy when the synchrotron radiation damping is very weak. The single bunch instability caused by the short-range wake field can be analyzed with the two particle model where head and tail particles each carry a charge of $n_b e/2$. The head particles disturb the EC distribution and the oscillation in the bunch head will be transferred to the bunch tail. For sufficiently long bunches, $\omega_e\sigma_z > c\pi/2$, the wake field felt by the tail particle is $W_{0,SB} \approx 8\pi\rho_e C/N_b$. $C$ is the circumference of the ring and $\rho_e$ is the volume density of the accumulated electron cloud. The single bunch instability manifests itself as strong-tail or transverse mode coupling instability (TMCI). With the strong head-tail model, the dimensionless parameter $\Gamma = \frac{N_b r_p W_{0,SB}\bar{\beta}}{16\gamma\nu_s} < 1$, is used to give the threshold of the wake field. The EC threshold density for the instability is expressed as $\rho_{e,threshold} < \frac{2\gamma\nu_s}{r_p\pi C\bar{\beta}}$. Rough estimates on TMCI and the density threshold for SPPC are summarized in Table 4. Some measures such as solenoid magnetic fields, clearing electrodes, or pipe coating should be taken to diminish the electron cloud.

The accumulated electron cloud as a focusing force on the proton beam will cause incoherent tune shift as the counterpart to space charge. Assume the EC is transversely uniform around the beam, then the tune shift is given by the formula [20]: $\Delta\nu = \frac{r_p}{\gamma}\bar{\beta}\rho_{ec}C$. A larger tune shift can lead to a severe drop in luminosity. For SPPC, with an average betatron function of about 100 m, the tune shift is estimated to be about 0.00225 which cannot be ignored when the EC density is about $1.0 \times 10^{13}$ m$^{-3}$. Therefore, in the lattice design, it will be necessary to consider the tune shift caused by the EC.

Because of very high synchrotron radiation power and low-temperature beam pipes for the superconducting magnets at SPPC, the deposited power on the beam screen from the secondary electron multipacting may be a serious issue. The measured deposited power in the dipole magnets of LHC has proven to increase exponentially to about 10 W/m, when SEY is larger than 1.4. Therefore, SEY at SPPC should be controlled to stay below 1.4 or even 1.2 by coating TiN or NEG on the internal walls of the vacuum chamber and devices inside the vacuum.

It has been noted [23] that the central electron cloud density in quadrupoles can be 2-3 orders of magnitude higher than in dipoles. This puts a severe constraint on the SEY in quadrupoles, if serious effects are to be avoided. A solution to this problem would be to add dipole fields to the focusing elements, making them combined function magnets. The dipole field needs only to be strong enough to move the zero field axis until it is outside the beam screen. This option would increase the relative lengths of the focusing elements, but probably not change the overall magnet length of bends and the focal distance.

Most parameters in Table 4 are hardly changed if the bunch spacing is reduced, assuming that the average current is maintained: $n_b/L_{sep}$ = constant. However, as the bunch spacing is reduced, the parameter $n$ changes rapidly. For a bunch spacing of 5 ns $n >> 1$ which should not be a problem; a large $n$ corresponds to an almost electrostatic field that can support an electron cloud, but does not amplify it by multipacting. As mentioned above, the cloud will depend only on its initial population from photo emission. With the slotted beam screen of figure 7b, this should not be a problem. However, an intermediate bunch spacing, $n \sim 1$, is the resonant case of maximal growth.

## 2.5 Beam loss and collimation

### 2.5.1 Beam loss

Beam losses will be extremely important for safe operation in a machine like SPPC where the stored beam energy will be 6.6 GJ per beam. Beam losses can be divided into two classes, irregular and regular [24-25]. Irregular beam losses are avoidable losses and are often the result of a misaligned beam or due to a fault in an accelerator element. A typical example is a trip of the RF, which causes loss of synchronization during acceleration and collisions. Vacuum problems also fall into this category. Such losses can be distributed around the machine. A well designed collimator system might collect most of the lost particles, but even a fraction of the lost particles may cause problems at other locations. Regular losses are non-avoidable and localized in the collimator system or on other aperture limits. They will occur continuously during operation and correspond to the lifetime and transport efficiency of the beam in the accelerator. The lowest possible losses are set by various effects, e.g. Touschek effect, beam-beam interactions, collisions, transverse and longitudinal diffusion, residual gas scattering, halo scraping and instabilities [25].

**1) Touschek effect:** This, also referred to as intra-beam scattering, is caused by the scattering of charged particles within an individual bunch, and their subsequent loss. It is typically estimated by an average of the scattering rate around the ring [26].

**2) Beam-beam interactions:** Beam-beam interactions at the IPs produce collisions for physics experiments, but also elastic and inelastic scattering that will lead to emittance blow-up and beam loss [26-27].

**3) Transverse and longitudinal diffusion:** Resonance crossings or unstable motion caused by unavoidable field errors and higher order multipoles can cause beam particles to leave their trajectories and strike the machine aperture. Particles inside the dynamic aperture may also diffuse out from the core of the beam and into the unstable region, e.g. through intra-beam scattering, beam-gas scattering and beam-gas bremsstrahlung [26, 28].

**4) Residual gas scattering:** This includes inelastic beam-gas nuclear inelastic interactions (both quasi-elastic and diffractive), elastic beam-gas nuclear elastic interactions (both coherent and incoherent), and Coulomb scattering. These effects degrade the beam quality and can also cause immediate beam loss [25, 27].

**5) Collimator tails:** Collimation is done in both betatron and momentum cleaning insertions. Protons that pass close to, or are only partially stopped by the collimators, can be deflected, and must be intercepted by tertiary and even quaternary collimators. But there is always some inefficiency in these systems leaving tails, also known as "tertiary/ quaternary beam halo" that can be lost in other locations in the ring [25, 29].

**6) Instabilities:** A beam becomes unstable when the moments of its distribution exhibit exponential growth (e.g. barycenters and standard deviations in different coordinates) which result in beam loss or emittance growth. There are a wide variety of mechanisms which may produce collective beam instabilities, with the most important ones being the electron cloud effect as described above and coupling impedance.

### 2.5.2 Collimation

For high-power proton accelerators, halo particles might potentially impinge on the vacuum chambers and get lost. The radiation from the lost particles will trigger quenching of the superconducting magnets, generate unacceptable background in detectors, damage radiation-sensitive devices, and cause residual radioactivity that prevents hands-on maintenance. These problems can be addressed by collimation systems which confine the particle losses to specified locations where better shielding and heat-load transfer are provided. For high-energy proton-proton colliders with very high stored energy in the beams, like SPPC, the situation is even more complicated, mainly because extremely high collimation efficiency is required. In addition, it is very difficult to collimate very high energy protons efficiently.

To illustrate the likely systems needed for the SPPC, we discuss first those used successfully in the LHC, even though it has lower beam energy and stored energy. The LHC uses 98 two-sided and 2 one-sided movable collimators, for a total of 396 degrees of freedom, which provide a four-stage collimation system to tackle 100 MJ of stored energy per beam [30-31]. LHC is now upgrading the systems for future operation at their design energy of 14 TeV (Center of Mass), and will do additional improvements for the high-luminosity upgrade (HL-LHC) [32]. Two warm interaction regions (IRs) or long straight sections are used to provide betatron collimation and momentum collimation. Both collimations employ the sophisticated multi-stage collimation method, and the main difference between the two is that a modest dispersion function is introduced in the long straight, which is required for the momentum collimation but there is no such need for the betatron collimation.

With the multi-stage collimation method [33], the primary collimators of small thickness are the closest to the beam in transverse phase space and will scatter the primary halo particles. They must be located at large $\beta$ value to maximize the impact parameters and reduce the out-scattering probability. The secondary and sometime even tertiary collimators will intercept and stop part of the scattered particles; however, they also produce out-scattered particles, which are called secondary and tertiary beam halos. The absorbers will stop the showers from upstream collimators and the additional tertiary or quaternary collimators are used to protect the superconducting quadrupole triplets at the colliding interaction regions directly [31]. The introduction of the collimation system not only demands precious space in the rings, but also increases the coupling impedance, important for collective beam instabilities.

For SPPC, the stored energy in the beam is as high as 6.6 GJ per beam, about 16 times that of the LHC at design energy. Therefore, for the same beam loss power, and to prevent frequent SC magnet quenching, the cleaning inefficiency at SPPC should be about 1/16 of that at the LHC. This means a cleaning inefficiency of only $4.3 \times 10^{-6}$. Five-stage collimation systems for both betatron and momentum collimations are foreseen. Fig. 4 shows the schematic for a five-stage collimation system. Two long straight sections of about 850 m provide the required space for hosting the collimation systems. The one for momentum collimation should be designed to have modest dispersion functions.

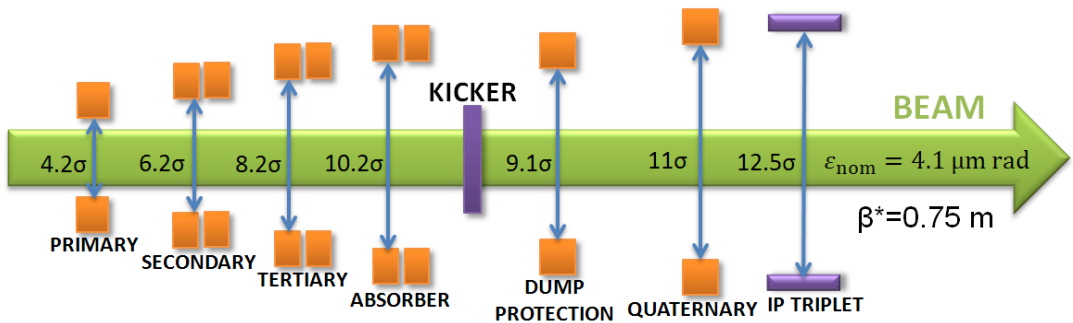

Fig. 4: Schematic for the multi-stage collimation system at SPPC

Besides the method used at LHC, other novel methods will be considered, including the one studied in CERN and FNAL with bent crystals [34-35], and the one employing nonlinear magnets to enhance the collimation efficiency [36-37]

# 3 Key technical systems

## 3.1 High-field superconducting magnets

### 3.1.1 Requirements of the high-field magnets for SPPC

To bend and focus the high energy proton beams, SPPC needs thousands of high-field dipoles and quadrupoles installed around a tunnel 54.4 km in circumference. The nominal aperture in these magnets is 50 mm. The field strength of the main dipoles is 20 T. A field uniformity of $10^{-4}$ should be attained up to 2/3 of the aperture radius. The magnets are designed to have two beam apertures of opposite magnetic polarity within the same yoke (2-in-1) to save space and cost. The currently assumed distance between the two apertures in the main dipoles is 330 mm, but this could be changed based on detailed design optimization to control cross-talk effect between the two apertures, and with consideration of overall magnet size. The outer diameter of the main dipole and quadrupole magnets should not be larger than 800 mm, so that they can be placed inside cryostats having an outer diameter of 1400 mm. The total magnetic length of the main dipole magnets is about 39 km out of the total circumference of 54.4 km. If the length of each dipole magnet is 15 m, then about 2500 dipole magnets are required. High gradient quadrupoles for SPPC are divided into the following three groups:

1) those at the IPs with single aperture, diameter D = 60 mm, and pole-tip field $B_{pole}$ = 20 T;
2) those in the matching section, D = 60 mm, $B_{pole}$ = 16 T;
3) those in the arcs, D = 50 mm, $B_{pole}$=16 T.

The ones in the matching sections and arcs are 2-in-1 yoke-sharing magnets.

### 3.1.2 Current status of high-field accelerator magnet technology

One of the most challenging technologies for SPPC is the development of the high field superconducting magnets. All the superconducting magnets used in present accelerators are made with NbTi. These magnets work at significantly lower field than the required 20 T (23.5 T is really required to have an operational margin), e.g., 3.5 T at 4.2 K at RHIC and 8.3 T at 1.8 K at LHC. As shown in Fig. 5, the critical "engineering" current density $J_E$ of most superconductor wires falls rapidly with the magnetic field. A reasonable design of accelerator magnets requires that the average $J_E$ of the cable should be above 500 A/mm$^2$ at the desired field. This criterion suggests that it should be possible to develop a dipole with Nb$_3$Sn of 15-16 T, but for 20 T one has to look for alternate superconductors. Fortunately, the advent of High Temperature Superconductors (HTS), whose current carrying capacity decreases only slowly with field (see Fig. 5), should allow magnets with much higher magnetic fields. It appears reasonable to build dipoles with fields of 20 T, using NbTi and Nb$_3$Sn coils combined or Nb$_3$Sn coils alone to provide a field of 15 T, together with 5 T provided by HTS (Bi-2212 or ReBCO) insert coils.

Development of superconducting dipole magnets started more than thirty years ago in US laboratories, as shown in Fig. 6. At BNL the Sampson magnet obtained 5-T main field in the late 1970's, which was followed by LBNL-D10 and CERN-Asner that reached 8-9 T in the late 1980's. The Twente-MSUT Nb$_3$Sn magnet was the first dipole magnet with a field beyond the limitation of Nb-Ti. LBNL has held dipole magnet records for the past fifteen years: Their D20 dipole reached 13.5 T in a 50-mm aperture in 1997; HD2 dipole reached 13.84 T in an aperture of about 40 mm in 2007; HD1a dipole reached a peak field of 15.4 T but without an accelerator aperture or appropriate field quality [38]; RD3C dipole reached 10 T in a 35-mm twin aperture with a common coil configuration. All these magnets were fabricated based on "Wind and React" technology and tested at 4.5 K. A similar common coil magnet was developed using "React and Wind" technology at BNL and reached over 10.2 T with a 31-mm aperture. All of these were R&D magnets and the current maximum dipole field in a real accelerator remains the LHC dipole's 8.3 T. To raise it to 20 T in 15 years or about by the year 2030 will require significant R&D in developing both the superconductor technology and the magnet technology.

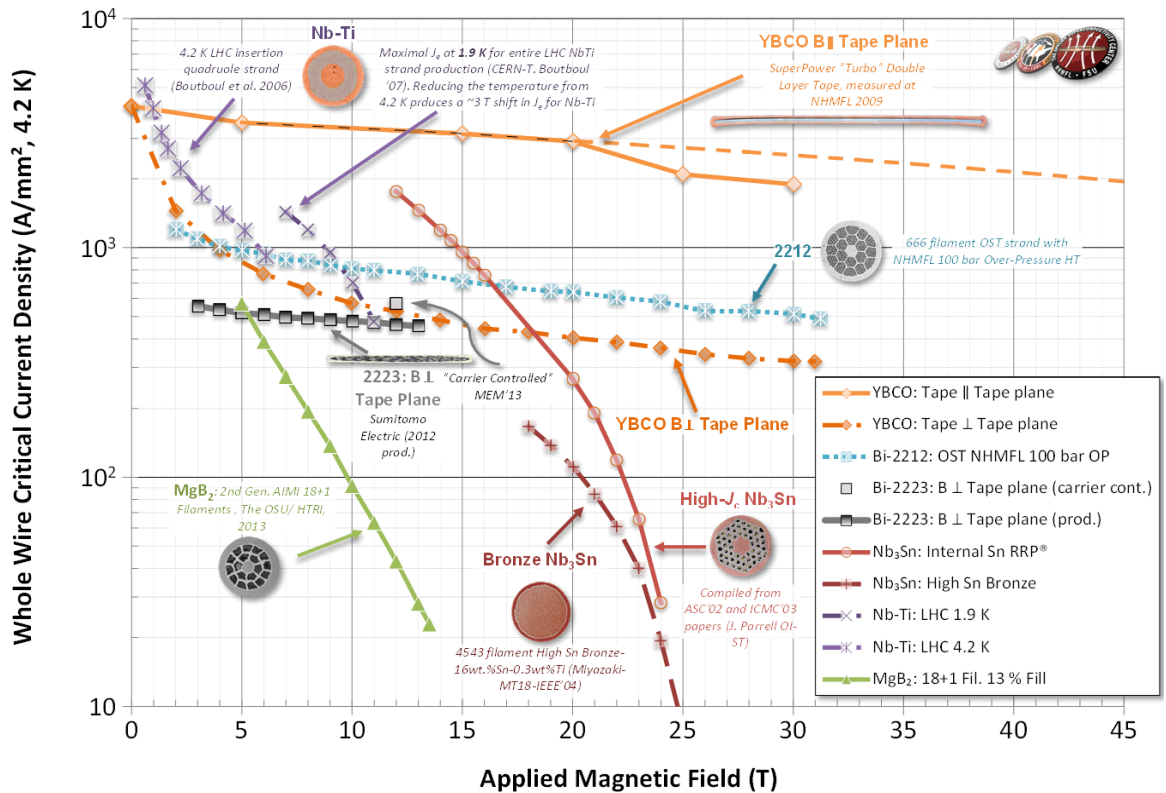

Fig. 5: Whole wire critical current density of main superconductors at 4.2 K [39]

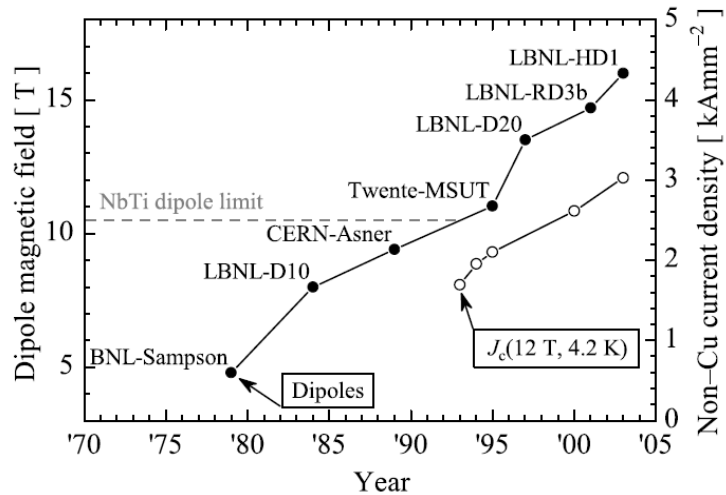

Fig. 6: Evolution in the highest field in Nb$_3$Sn dipoles [40]

### 3.1.3 Challenges to meet the SPPC requirement

1) *Obtaining the required performance, volume of production, and cost of superconductors will be a challenge*: Thousands of tons of $Nb_3Sn$ and HTS superconductors will be needed. The cost for the superconductor materials is likely to be a cost driver. Potential further increase of $J_E$ in both $Nb_3Sn$ and Bi-2212 [41-42] may be expected, and would reduce the required quantity of superconducting materials and the cost of the SPPC project. It will be a major challenge for superconductor manufacturing industries to improve the performance, reduce the cost, and scale up for the volume of superconductors required for the project.

2) *Constraining the high magnetic forces at 20 T*: The magnetic force in superconducting coils increases as the square of the field. If not managed [43], for 20-T, the stress in $Nb_3Sn$ or HTS coils will be above 200 MPa. As both $Nb_3Sn$ and Bi-2212 superconducting materials are strain sensitive, with $J_E$ going down quickly with increasing strain, some management will be required. ReBCO, on the other hand, can tolerate much higher stress and strain (a factor of three more) without showing any degradation, but the magnetization in ReBCO is more severe than for Bi-2212.

3) *Reducing training*: Training requiring multiple quenches, before reaching the desired magnet field, is probably not acceptable in such a program, because of the expense of cryogenic power. Its reduction, or elimination, has been shown to require very good support with significant pre-compression.

4) *Achieving the required field quality with HTS coils, particularly those wound with ReBCO tape*: The current distribution within a filament or tape depends on the history of fields it has seen. This 'magnetization' depends strongly on the dimensions of the individual conducting strands. Finer strands give much less magnetization. LTS (Low temperature superconductor) conductors such as NbTi are made of thousands of small filaments with diameter of only a few microns. The filaments in current Bi-2212 conductors are larger than those in NbTi, and the ReBCO tape is a single 'filament' and is orders of magnitude larger. This will make it difficult for the magnets with the HTS coils to reach the field uniformity level of $10^{-4}$ with the present designs. Some innovative solutions are being studied.

5) *Achieving quench protection of HTS coils*: The quench propagation speed in HTS coils is hundreds of times lower than in LTS coils. This makes the present quench detection and protection methods unsuitable for HTS coils. Innovative solutions are being studied.

6) *Developing twin aperture 20-T magnets in an outer diameter of 800 mm*: The magnetic cross-talk between the two apertures should be controlled without increasing the size of the magnet. Moreover, the iron saturation effect should be carefully controlled to attain field quality of $10^{-4}$ at both injection (low current) and collision (high current) fields.

### 3.1.4 Preliminary design for the SPPC superconducting magnets [44]

A preliminary conceptual design of a 2-in-1 common coil dipole of 50 mm in aperture and 20 T in field is shown in Fig. 7. The design is based on the current $J_E$ level of the superconductors at 4.2 K. The large bend radius at the common coil ends allows the use of "React and Wind" technology for coil fabrication. The short sample dipole field of the magnet is 22 T at 4.2 K (the figure shows a 20-T dipole field at 91% load line ratio). The outer diameter of the iron yoke is 720 mm. Six racetrack coils are needed to reach a short sample field of 22 T. Two inner coils are made with Bi-2212 and four outers with Nb₃Sn. All the coils have simple racetrack geometry except for a small one with a few turns at the pole. The Bi-2212 coils are wound with 20-mm wide cables. The cable has fifty Bi-2212 round wires of 0.8 mm in diameter. The outer four Nb₃Sn coils are wound with two types of cable: 22-mm width wider cable fabricated with fifty-five Nb₃Sn wires and 15-mm width narrow cable fabricated with thirty-seven Nb₃Sn wires. Both operate at the same current of 14.5 kA at 20 T, providing 'graded' current densities in different field regions. The critical current density of the Nb₃Sn and Bi-2212 superconductors is calculated with the data in [39-40]. The Bi-2212 data was with reaction in with 100 bar overpressure, something probably more practical when applied before winding, than after winding in a 15-m long magnet.

For a 20-T common coil test dipole of 1 m length, the required length for the 0.8-mm diameter Nb₃Sn wire is 39 km (about 166 kg) and for the 0.8-mm diameter Bi-2212 wire is 13.8 km (about 60 kg). ReBCO wires will also be considered in future design studies.

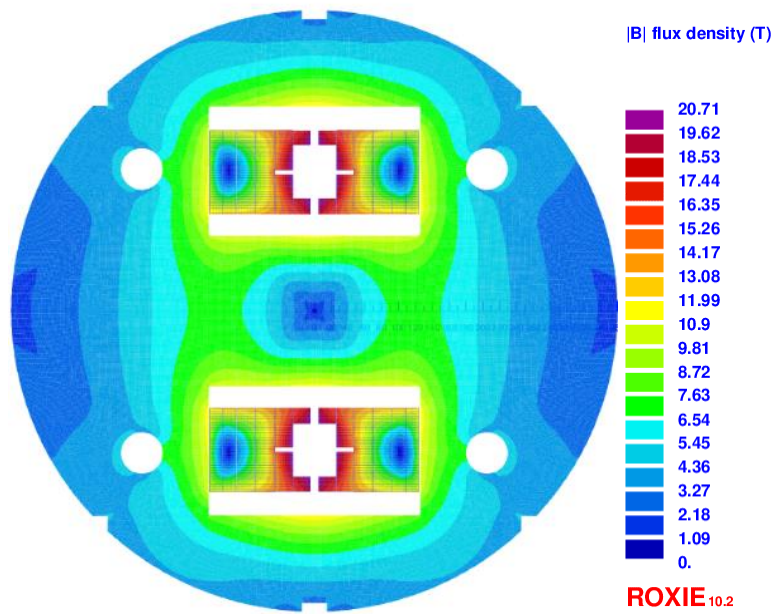

Fig. 7: Conceptual design of the 20-T dipole for SPPC

To address the problem of high synchrotron radiation load in the low-temperature vacuum pipe, an open mid-plane structure can also be considered.

Operation at 1.8 K, instead of 4.2 K, is another option worth study. The quantities of NbTi and Nb₃Sn, and their cost, would be reduced, but the cryogenic cost would be greater. An optimization is required.

## 3.2 Vacuum and beam screen

### 3.2.1 General vacuum considerations

SPPC has three vacuum systems: Insulation vacuum for the cryogenic system; beam vacuum for the low-temperature sections; and beam vacuum for the chambers in the room-temperature sections.

*1) Insulation vacuum*

The aim here is only to avoid convective heat transfer and there is no need for high vacuum. The room-temperature pressure in the cryostats before cool-down does not have to be better than 10 Pa. Then, so long as there is no significant leak, the pressure will stabilize around $10^{-4}$ Pa, when cold. As a huge volume of insulation vacuum is needed at SPPC, careful design is needed to reduce the cost.

*2) Beam vacuum in cold sections*

In interaction regions or around experiments where superconducting quadrupoles are used, the vacuum has to be very good (less than $10^{13}$ $H_2$ per $m^3$) to avoid creating background in the detectors. But the beams are straight here and there is relatively little synchrotron radiation.

In the arcs, the requirement is based on the beam lifetime, which depends on the nuclear scattering of protons on the residual gas [5]. To ensure a beam lifetime of about 100 hours, the equivalent hydrogen gas density should be below $10^{15}$ $H_2$ per $m^3$. The problem here is the huge synchrotron radiation power. If allowed to fall directly on the magnet bore at the magnet temperature of 4.2 K, the wall power needed to remove it would be grossly too high. It has to be intercepted on a beam screen, which works at a higher temperature, e.g. 40-60 K and is located between the beam and cold bore (see below). This screen, at such a temperature, will desorb hydrogen gas, particularly if it is directly exposed to synchrotron radiation. The space outside the screen will be cryopumped by the low temperature of the bore. Slots must be introduced in the shield to pump the beam space. However, with the core at 4.2 K, the pumping speed of $H_2$ is low, thus one may need to use other auxiliary methods, such as cryosorbers used at LHC [45].

*3) Beam vacuum in warm sections*

The warm regions are used to house the beam collimation, injection, and extraction systems. They use warm magnets to avoid superconductor quenching from the inevitable beam losses in these locations. They have difficult vacuum pumping requirements due to desorption from the beam losses. Non Evaporable Getter (NEG) is probably required. At least these sections are of limited overall length.

*4) Vacuum instability*

Vacuum instability issues need further investigation [46].

### 3.2.2 Beam screen

The main function of a beam screen is to shield the cold bore of the superconducting magnets from Synchrotron Radiation (SR) [47]. At SPPC, synchrotron radiation is especially strong because of the very high beam energy and high magnetic field in the arc dipoles. The estimated SR power is about 57 W/m per aperture in the arc dipoles. This is much higher than the 0.22 W/m at LHC [48], and greatly increases the difficulty of the beam screen design.

The operating temperature of the screen must be high enough to avoid excessive wall power needed to remove the heat. But not too high to avoid excessive resistivity of the copper coating on its inside surfaces, leading to excessive impedance, and to avoid radiating too much power on to the bore at 4.2 K.

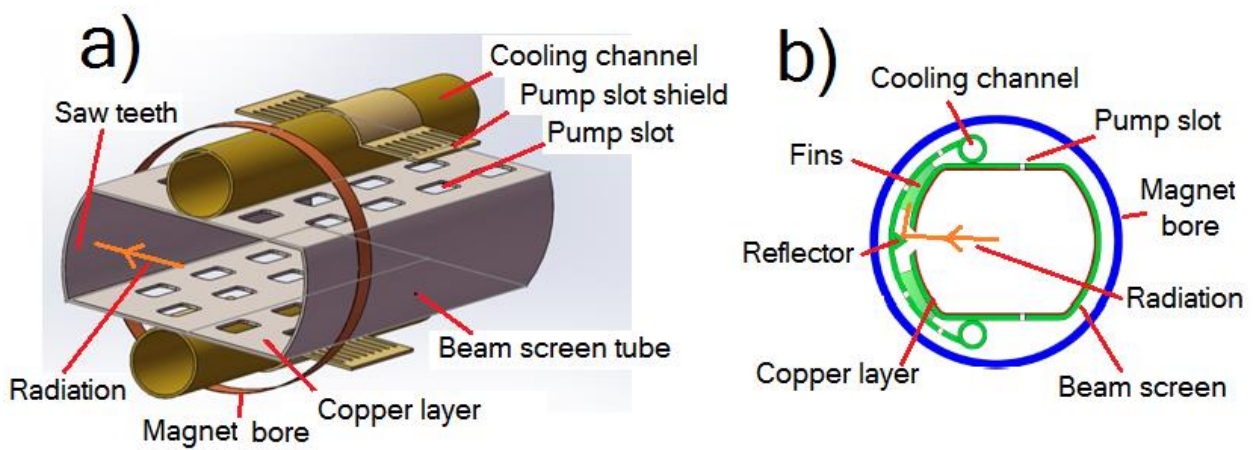

Fig. 7: Schematic for a beam screens: a) under consideration at SPPC; b) As proposed for FCC

The design must satisfy requirements of vacuum stability, mechanical support, influence on beam dynamics and refrigeration power. Fig. 7a shows a schematic for a basic beam screen under consideration at SPPC. Fig. 7b shows an alternative screen design as discussed [13].

The main challenges are:

*1) Synchrotron Radiation*

The SR power deposition at the SPPC main dipoles is two orders higher than that at LHC [48]. If absorbed at 4.2 K, the cryogenic load would be excessive and very expensive, so a beam screen between the beam and cold bore is essential. The operating temperature of the beam screen should be high for wall power economy, and to decrease technical difficulty, but should not be too high. The inside of the screen is coated with a copper layer to reduce the resistive impedance. At higher temperatures this impedance, from its higher electrical resistivity, will be increased, leading to worse collective beam instability. The operating temperature is also constrained to limit heat radiation and conduction to the cold bore, and by considerations of desorption. The operating temperature should be chosen carefully. Different refrigerants can be considered, such as liquid neon or liquid oxygen.

*2) Electron cloud*

A proper beam screen structure can restrain the generation of photo-electrons feeding an electron cloud. In the basic screen design (Fig.7a), the synchrotron radiation falls directly on a saw-tooth shaped surface on the inner wall of the screen. A low desorption coefficient is needed to minimize electron emission. Primary emission on the mid-plane is trapped by the dipole magnetic field. But forward scattering of the SR can reach azimuths where the magnetic field does not help.

The proposed FCC-hh design (Fig. 7b) has a slit in the outer mid-plane of the screen, and 45° surfaces that reflect the synchrotron radiation up or down into confined absorption structures where photo-desorption is not a problem.

In either case, the inner screen's surface should be coated by a thin film of low secondary electron yield to reduce electron production.

3)  *Vacuum*
Vacuum in the beam screen will depend on several factors: the beam structure, the beam energy, the beam population, the critical photon energy and synchrotron radiation power. Beam structure has an important effect on the buildup of the electron and ion clouds which may lead to vacuum instability. Pumping speed is the dominant factor for vacuum stability. The beam screen must be designed with sufficient transparency to retain an effective pumping speed. However, good transparency obtained by adding more slots will increase the resistive impedance which may cause beam instabilities.

4)  *Magnet quenches*
The beam screen should have sufficient strength to resist the pulsed electromagnetic forces generated by a superconducting magnet quench [49]. Stainless steel can also be used as the base structure material, reducing such forces, but a thick copper film of 75 µm coated on the base to decrease the wall impedance produces a strong source of electromagnetic force. The thinner the film, the smaller the force, but the higher the resistive impedance.

5)  *Impedance*
The shape and size of the beam screen structure needs to be optimized in order to decrease the transverse wall impedance.

6)  *An ideal solution?*
An ideal design might separate the two functions of the beam screen: The screen itself (on the right in fig 7b), with the slot on the outer side would be run at a relatively lower temperature to control the impedance, while the absorption structures (on the left in Fig. 7b) would be at a higher temperature to minimize the wall power needed to extract the synchrotron radiation power. The greatest challenge may be to restrain the forces on the screen itself while minimizing thermal contact with the warmer absorption structures.

## 3.3 Other technical challenges

Besides the two key technologies described above, high-field magnets and vacuum/beam screens, there are other important technologies requiring development in the coming decade in order to build SPPC. Among them are the machine protection system that requires extremely high efficiency collimation, and a very reliable beam abort system. These are important for dumping the huge energy stored in the circulating beams, when a magnet quenches, or another abnormal operating condition occurs. If the extraction system has to be installed in a relatively short straight section, one has to develop more powerful kickers.

A complicated feedback system is required to maintain beam stability. The beam control system also controls emittance blow-up in the main ring which is important for controlling beam-beam induced instabilities and for leveling the integrated luminosity.

Beam loss control and collimation in the high-power accelerators of the injector chain pose additional challenges. A proton RCS of 10 GeV and a few MW is still new to the community, and needs special care. The gigantic cryogenic system for magnets, beam screens and RF cavities also needs serious consideration.

# 4 Configuration of the accelerator complex

## 4.1 Injector chain

The injector chain by itself is an extremely large accelerator complex. To reach the beam energy of 2.1 TeV required for the injection into the SPPC, we require a four-stage acceleration system, with energy gains per stage between 8 and 18. It not only accelerates the beam to the energy for injection into the SPPC, but also prepares the beam with the required properties such as the bunch current, bunch structure, and emittance, as well as the beam fill period.

The four stages are shown in Fig. 8, with some more parameters given in Table 5. The lower stage is, the higher repetition rate it has. The p-Linac is a superconducting linac with a repetition rate of 50 Hz. The p-RCS is a rapid cycling synchrotron with a repetition rate of 25 Hz. The MSS has a relatively lower repetition rate of 0.5 Hz. The SS which is based on superconducting magnets with maximum dipole field of about 8 Tesla is even slower. The higher repetition rates for the earlier stages help reduce the SS cycling period and thus the overall SPPC beam fill time. For easier maintenance and cost efficiency, as well as the physics programs, the first three stages will be built in a relatively shallow underground level, e.g., -15 m, whereas the SS with a much larger circumference will be built in the same level as the SPPC or about -100 m.

As shown in Table 5, for the SPPC, the different stages are needed for only fractions of the time. They could operate with longer duty cycle, or continuously, to provide high-power beams for other research applications, when not used for the SPPC. As the present bunch population at the SPPC is limited mainly by the SR power, the accelerators of the injector chain have the potential to load more accumulated particles in a pulse or deliver higher beam power for their own diverse applications when not serving the SPPC.

For such a complex injector system, it will take about 10 years to build and commission stage-by-stage. Thus hopefully the construction of the injector accelerators can be started several years earlier than the SPPC, and this means that it overlaps with the CEPC physics operation.

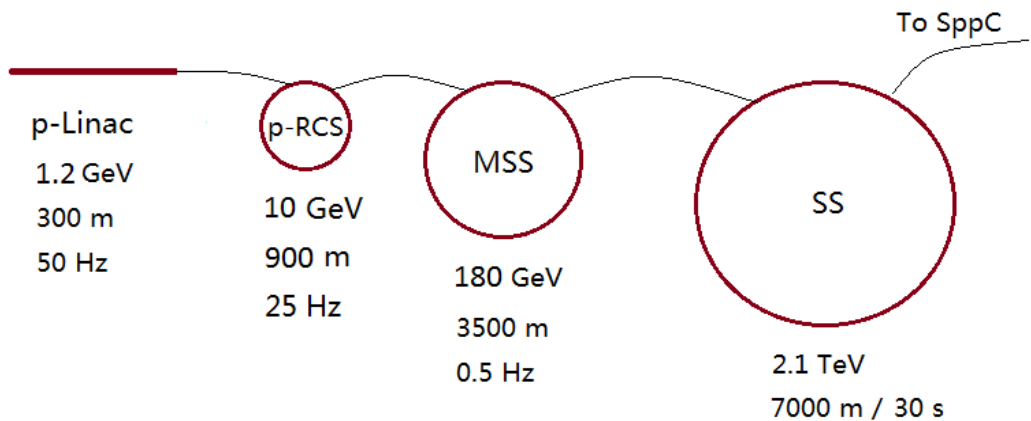

Fig. 8: Injector chain for the SPPC

*Linac*

Superconducting linacs have undergone tremendous development [50] and will presumably make even more progress in the next decade. Hence we have adopted a 1.2 GeV in energy and 50 Hz in repetition rate for the p-Linac. The continuous beam power is 0.84 MW. At least half of this could be available for other applications.

*Rapid Cycling Synchrotron (p-RCS)*

The continuous beam power from p-RCS is 3.5 MW. Only one other proton driver study (for a future Neutrino Factory) has performance close to this [51]. The high repetition rate of 25 Hz will shorten the beam filling time in the MSS. Only a fraction of this power is needed to fill the MSS. Thus most of the beam pulses from the p-RCS could be used for other physics programs. The p-RCS will use mature accelerator technology but be on a larger scale than existing rapid-cycling proton synchrotrons.

*Medium Stage Synchrotron (MSS)*

The MSS has beam power similar to the p-RCS but with much higher beam energy and much lower repetition rate. The SPS at CERN and the Main Injector at Fermilab are two good examples for its design. But due to higher beam power, the beam loss rate must be more strictly controlled. A bunch splitting technique using a multiple harmonic RF system is used here, and in the SS, to prepare the bunch gap of 25 ns or less, as required by the SPPC. Certainly, the beam from the MSS will find additional physics programs other than only being the injector for the SS.

*Super Synchrotron (SS)*

The SS will use superconducting magnets similar to those used at the LHC, but with a higher ramping rate. Here, we do not need to consider synchrotron radiation because of the much lower energy. There are no apparent critical technical risks in building the SS. It is unclear if the beam from the SS can find its own physics programs besides being the SPPC injector.

Table 5: Main parameters for the injector chain at SPPC

| | Energy | Average current | Length/ Circum. | Repetition Rate | Max. beam power | Dipole field | Duty factor for next stage |
|---|---|---|---|---|---|---|---|
| | GeV | mA | M | Hz | MW/MJ | T | % |
| p-Linac | 1.2 | 1.4 | ~300 | 50 | 1.6 | - | 50 |
| p-RCS | 10 | 0.34 | 900 | 25 | 3.4 | 1.0 | 6 |
| MSS | 180 | 0.02 | 3500 | 0.5 | 3.7 | 1.4 | 13.3 |
| SS | 2100 | - | 7000 | 1/30 | 90 | 8.0 | 1.0 |

A dedicated heavy-ion linac (I-Linac) together with a new heavy-ion synchrotron (I-RCS), in parallel to the proton linac/RCS, is needed to provide heavy-ion beams at the injection energy of the MSS, with a beam rigidity of about 36 Tm which is the same as the 10 GeV proton beam.

## 4.2 Integration of CEPC and SPPC

The present proposal calls for continuing the CEPC *e+e-* program after SPPC is brought into operation. Housing both CEPC and SPPC in a common underground tunnel and operating them alternatively, or simultaneously, would be unprecedented. There is also a plan for making ep and eA collisions using the CEPC electron beam and one SPPC beam. While in principle it is plausible, there are technical and operational risks. Therefore we must plan for such operation at an early stage of the CEPC-SPPC project. In this section, we first present a brief discussion on the anticipated risk factors and suggestions for mitigating these risks. We then address several special issues for achieving good integration of the two facilities.

### 4.2.1 Project Uncertainty

While it is necessary, and also advantageous, to start the basic planning and preliminary conceptual design studies for SPPC at the present time, nevertheless, there are many intrinsic uncertainties which could prevent us reaching our goals. The first and perhaps the greatest challenge is anticipating the long term science priorities, bearing in mind that the project life cycle of CEPC-SPPC could easily exceed 40 years. The development of science may change the research goals. In addition, accelerator technology will surely advance in key areas, such as ultra-high-field superconducting magnets, in ways we cannot now predict. There are cases of projects that failed to reach important science goals due to various limits or constraints posed in an early phase of the projects. To mitigate these risks and improve the chance of success of SPPC, one should try, within the foreseeable budget scenarios, to leave large margins in the technical specifications of the facility. This includes maximally expanding the SPPC performance range (primarily the energy and luminosity), and to take the least optimistic forecast of technology developments. An increased circumference of the main tunnel could be an example of this. It would reduce the synchrotron power in CPEC, lower the required magnetic fields in SPPC, and leave future options for energy and luminosity upgrades.

**4.2.2 Geometric Constraints and Considerations**

When completed, the CPEC and SPPC will, in the same tunnel, have three collider rings: two for the proton beams, one shared by the electron and positron beams, plus a full-energy lepton booster ring. The later addition of a full circumference accumulator ring for protons has also been suggested. Space between the rings will be needed for machine maintenance. This will require a sufficiently large tunnel cross section. Detectors for the two colliders will occupy different straight sections of the rings, but by-passes of the detectors, for the non-intersecting beams, will be required; at these energies, this will not be trivial. At other locations where large machine elements, such as SRF modules, are installed, the other beam lines must be kept sufficiently apart.

**a)    Construction and Commissioning Considerations**

Installation of SPPC, after CEPC is operating, will pose one of the biggest challenges. It will probably require a long (multi-year) shutdown of CEPC, affecting its physics program. To avoid too long a shutdown, the LHeC (a Large Hadron-electron Collider envisioned at CERN) chooses a linac-ring collider scenario, over a ring-ring one. The SPPC cannot do this, so the design should be optimized to enable rapid installation and commissioning to minimize the CEPC shutdown. Protection of the CEPC machine during the construction and commissioning of the SPPC will be challenging.

**b)    Operational Considerations**

Placing the CEPC and SPPC collider rings side-by-side may provide an opportunity for sharing resources and equipment such as the liquid helium supply line and power supply line and network communication lines, leading to a cost reduction for SPPC. Radiation protection may also be shared, requiring less or no upgrade for operating the SPPC, particularly under the operational mode of alternative running of the two colliders. The central control system and machine control center staffing may also be shared. By having these two installations at the same site other cost savings will be the shared campus with its infrastructure such as administration, on-site computers, user amenities and library.

There will also be challenges in the simultaneous operation of the two super colliders. There will be considerable load variations on the power grid, as CEPC enters the top-off mode, and when the SPPC hadron injector complex (including the linac and three booster rings) prepares and injects a proton beam into the collider storage ring. The CEPC-SPPC infrastructure must be designed to handle such load capacity variations.

Large temperature variation may affect proper functioning of some electronic systems. Heat generated from the machine elements of the two colliders must be removed efficiently, so as to maintain a steady temperature inside the tunnel.

Machine protection is another challenge for the CEPC-SPPC joint facility.  An event, or a major accident in one collider, could damage equipment in the other.

It is understood that simultaneous operation of two colliders will introduce an overhead and reduce the duty factors of an individual collider. Maintenance and repair of one collider may force suspension of operation and data-taking of the other collider. The ep and eA collision modes have less such problems.

## Acknowledgements

The authors would like to thank all the international experts who have provided us with useful comments during the reviews and on other occasions. We also want to thank the Directors of the participating Chinese laboratories for their great support of this study, and the US authors want to thank DoE for the support. The work is partially supported by National Natural Science Foundation of China (Project 11235012).